\documentclass{nature}
\usepackage[utf8]{inputenc}
\usepackage{tabulary,amsmath,amsfonts,amssymb}
\usepackage{url,multirow,cancel,tfrupee}
\makeatletter
\AtBeginDocument{\@ifpackageloaded{textcomp}{}{\usepackage{textcomp}}}
\makeatother
\usepackage{colortbl}
\usepackage{xcolor}
\usepackage{pifont}
\usepackage{graphicx} 
\usepackage{physics} 
\usepackage{float} 
\usepackage{epsfig,amssymb} 
\usepackage{epstopdf}
\usepackage[10pt]{moresize} 
\usepackage{subfig} 
\usepackage{wrapfig}
\usepackage[nointegrals]{wasysym}
\makeatletter

\begin{document}

\title{Quantum Simulation of Klein Gordon Equation and Observation of Klein Paradox in IBM Quantum Computer}
  
\author{
Manik Kapil$^{1}$\thanks{E-mail: manikkapil0@gmail.com}{ },
Bikash K. Behera$^{2}$\thanks{E-mail: bkb13ms061@iiserkol.ac.in }{ },
              Prasanta K. Panigrahi$^{2}$\thanks{Corresponding author.}\ \thanks{E-mail: pprasanta@iiserkol.ac.in}
              }
\maketitle 

\begin{affiliations}
     \item Department of Physics\unskip, 
    Atma Ram Sanatan Dharma College\unskip, Delhi University\unskip, Dhaula Kaun\unskip, 110021\unskip, New Delhi\unskip, India
    \item
      Department of Physical Sciences\unskip, 
    Indian Institute of Science Education and Research Kolkata\unskip, Mohanpur\unskip, 741246\unskip, West Bengal\unskip, 
    India
\end{affiliations}

\begin{abstract}
The Klein Gordon equation was the first attempt at unifying special relativity and quantum mechanics. While initially discarded this equation of ``many fathers" can be used in understanding spinless particles that consequently led to the discovery of pions and other subatomic particles. The equation leads to the development of Dirac equation and hence quantum field theory. It shows interesting quantum relativistic phenomena like Klein Paradox and ``Zitterbewegung", a rapid vibrating movement of quantum relativistic particles. The simulation of such quantum equations initially motivated Feynman to propose the idea of quantum computation. While many such simulations have been done till date in various physical setups, this is the first time a digital quantum simulation of Klein Gordon equation is proposed on IBM's quantum computer. Here we simulate the time-dependent Klein Gordon equation in a barrier potential and clearly observe the tunnelling of the particle and anti-particle through a strong potential claiming Klein Paradox. The simulation technique used here inspires the quantum computing community for further studying Klein Gordon equation and applying it to more complicated quantum mechanical systems.
\end{abstract}
    
\section{Introduction \label{qkg_Intro}}
The possibilities that quantum computation offers today in not only solving problems of quantum mechanics but also in various other algorithms is truly remarkable \cite{qkg_FarhiScience2001,qkg_HanCEC2000,qkg_ChuangNature1998,qkg_JonesNature1998,qkg_GuldeNature2003,qkg_HarrowPRL2009}. It is well known that Shor's algorithm \cite{qkg_ShorSIAM1997} once simulated on a quantum computer can be used to break our most common form of cryptography to secure data \cite{qkg_LomonacoarXiv2000}. Similarly many other algorithms like the Grover's algorithm \cite{qkg_GroverPRL1997} have also been shown to be exponentially faster than their classical counterparts. 

Since 1982 when Richard Feynman proposed the simulation of quantum systems using quantum computers \cite{qkg_FeynmanIJTP1982} there have been a major development on simulation of quantum mechanical systems. Quantum simulation has been found a great interest in various quantum systems including Hubbard model \cite{qkg_JakschPRL1998,qkg_DengWR2008}, spin models \cite{qkg_Garcia2004,qkg_LanyonSci2011}, quantum phase transitions \cite{qkg_Greiner2002,qkg_Pollet2010}, quantum chemistry \cite{qkg_Lidar1999}, quantum chaos \cite{qkg_Wien2002}, interferometry \cite{qkg_Leibfried2002,qkg_ViyuelanpjQI2018,qkg_LangfordnpjQI2017,qkg_AggarwalarXiv2018} and so on. Simulation of such quantum systems has shown to be more effective than those performed by classical systems \cite{qkg_BulutaScience2009}.

The simulation of quantum field theories has been performed by both analog quantum simulations, where the Hamiltonian is mapped to another system \cite{qkg_Gerretsma2010,qkg_Lamata2007}, or a digital quantum simulation where the Hamiltonian is split using the Suzuki Trotter formula \cite{qkg_Yosh1990,qkg_Sornborger1999,qkg_Hatano2005}, generally split into kinetic and potential energy terms \cite{qkg_BuchlerPRL2005,qkg_ZoharPRL2011,qkg_SzirmaiPRA2011,qkg_CiracPRL2010,qkg_MazzaNJP2012,qkg_KapitPRA2011,qkg_BermudezPRL2010,qkg_MaranerPLA2009,qkg_LeporiEPL2010,qkg_MaedaPRL2009,qkg_RappPRL2007,qkg_WeimerNAT2010,qkg_CasanovaPRL2011,qkg_DoucotPRB2004,qkg_LewensteinAP2007,qkg_JohanningJPB2009} and then simulated by appropriately decomposing the Hamiltonian into unitary operators and designing the quantum circuits on a quantum computer. Using the above technique, non-relativistic simulation of the Schrodinger equation \cite{qkg_Bogh1998,qkg_Giuliano2007}, relativistic simulations of  Dirac equation \cite{qkg_Gerretsma2010,qkg_Lamata2007} and the digital simulation of Dirac equation \cite{qkg_Gourdeau2017} have already been performed.

The Klein Gordon equation which can be derived from Schrodinger's equation was initially proposed for the unification of quantum mechanics and special relativity. This equation also shows "Zitterbewegung" and the Klein Paradox, where if the barrier potential V is greater than $mc^2$ the particle can easily penerate the barrier as if it wasn't there \cite{ qkg_DombeyPR1999,qkg_CalogeracosCP1999,qkg_CalogeracosIJMP1999}. Significant work has been done on the Klein Gordon equation including solving the equation \cite{qkg_Ebaid2009,qkg_Feshbach1958} and simulating it in classical systems \cite{qkg_Rusin2012}. This is the first time where we are simulating the Klein Gordon equation in a quantum system. It aims towards simulating the Klein Gordon equation for a single particle using digital quantum simulation. The basic technique used here can be utilized to effectively simulate the Klein Gordon equation that applies for any physical quantum system. Hence it illustrates an important application of quantum computers in the field of quantum simulation. This paper proposes a method for the simulation of the Klein Gordon equation and observes the so called Klein Paradox. We use the IBM quantum computer to simulate these equations using the QISkit, where a number of research works have been performed \cite{qkg_GarciaarXiv2017,qkg_DasarXiv2017,qkg_BKB1QIP2017,qkg_HBP17,qkg_MajumderarXiv2017,qkg_SisodiaQIP2017,qkg_BKB6arXiv2017,qkg_WoottonQST2017,qkg_BertaNJP2016,qkg_DeffnerHel2017,qkg_HuffmanPRA2017,qkg_AlsinaPRA2016,qkg_YalcinkayaPRA2017,qkg_GhosharXiv2017,qkg_KandalaNAT2017,qkg_Solano2arXiv2017,qkg_SchuldEPL2017,qkg_SisodiaPLA2017,qkg_1BKB6arXiv2018,qkg_2BKB6arXiv2018,qkg_3BKB6arXiv2018,qkg_4BKB6arXiv2018,qkg_5BKB6arXiv2018,qkg_6BKB6arXiv2018,qkg_Roffe2018,qkg_Plesa2018,qkg_ManabputraarXiv2018,qkg_RounakarXiv2018,qkg_SayanQIP2018}. We mainly use the local simulator device for performing the experiment and taking the results for the simulation purposes. 

\section{Results}
\subsection{Theoretical Protocol}
A common technique for solving the Klein Gordon equation classically is to represent the wavefunction as two different parts of a vector and then finding a Hamiltonian for that. This helps us to easily understand some of the peculiar natures of the Klein Gordon equation. We mainly use the form, $i\hbar\frac{\partial\varphi}{\partial t}=\hat{H}\varphi$, to solve the equation for the Hamiltonian $\hat{H}$ whose expression is given below. 

\begin{equation}
\hat{H}=\frac{\sigma{_3}+i\sigma{_2}}{2m}\hat{p}^2+\sigma{_3}mc^2+I \hat{V}
\end{equation}
where m, $\hat{p}$, V, c and I represent the mass of the particle, the momentum operator, the scalar potential, the speed of light and the identity matrix respectively where $\sigma{_i}$ (i = 1, 2, 3) are the Pauli matrices. The wavefunction $\varphi$ is a vector with two components.
\begin{equation*}
\varphi=
\begin{bmatrix}
    \phi \\
    \chi
\end{bmatrix}
\end{equation*}
After solving two simultaneous equations, we have 
\begin{equation}
\hat{H_{1}}\phi=\hat{K}(\phi+\chi)+\hat{V_{1}}\phi
\end{equation}
\begin{equation}
\hat{H_{2}}\chi=-\hat{K}(\phi+\chi)+\hat{V_{2}}\chi
\end{equation}
where
$\hat{K}=\frac{\hat{p}^2}{2m}$,
$\hat{V_{1}}=mc^2+\hat{V}$ and
$\hat{V_{2}}=-mc^2+\hat{V}$. 

The time evolution of a general Hamiltonian is given by
\begin{equation}
\ket{\psi(x,t+\delta)}=e^{-i\hat{H_{i}}t}\ket{\psi(x,t)}
\end{equation}
where $H_{i}=\hat{K}+\hat{V_{i}}$ and i=1,2.
Using the second order Suzuki Trotter decomposition \cite{qkg_Yosh1990,qkg_Sornborger1999,qkg_Hatano2005} we can decompose the unitary operators as follows.
\begin{equation}
e^{-i\hat{H_{1}}t}=e^{-i(\frac{\hat{p}^2}{2m}(\Box+\Box)+\hat{V}_{1})t} 
\label{qkg_Eq5}
\end{equation}
\begin{equation}
e^{-i\hat{H_{2}}t}=e^{i(\frac{\hat{p}^2}{2m}(\Box+\Box)-\hat{V}_{2})t}
\label{qkg_Eq6}
\end{equation}
In Eqs. \eqref{qkg_Eq5} \& \eqref{qkg_Eq6}, the kinetic energy portion can be solved by splitting the two momentum operators as given below. Here the box `$\Box$' denotes the component of the wavefunction used in the particular equation. 
\begin{equation}
e^{\pm i\frac{\hat{p}^2}{2m}(\Box+\Box)}=e^{\pm i\frac{\hat{p}^2}{2m}\Box}.e^{\pm i\frac{\hat{p}^2}{2m}\Box}
\end{equation}
 
These individual momentum operators can be expressed in momentum space by using quantum Fourier transformation \cite{qkg_Sornborger2012,qkg_Feng2013} and momentum eigenstates as shown in Eq. \eqref{qkg_Eq12}.
\begin{equation}
e^{\pm i\frac{\hat{p_{x}}^2}{2m}(\Box+\Box)}=F^{-1}e^{\pm i\frac{\hat{p_{p}}^2}{2m}\Box}.F.F^{-1}.e^{\pm i\frac{\hat{p_{p}}^2}{2m}\Box}F
\label{qkg_Eq12}
\end{equation}
where $\hat{p_{p}}$, F and $F^{\dagger}$ represent the momentum eigenvalues, Fourier transformation and it's inverse respectively. These operators can be implemented on a quantum computer by using a set of controlled phase gates and Hadamard gates \cite{qkg_Copper2002,qkg_AggarwalarXiv2018}.
The potential energy term for a double potential can be written as,
\begin{equation}
e^{-iV\hat{X}t}=I \otimes e^{-iV_{0}\sigma_{z}t}
\end{equation}

To express the equation as a digital quantum circuit we take the mass of the system as m=0.5 and use the co-ordinates system such that
$h=\sqrt{2}c=1$.
Using the above values, we have the full expressions for the unitary operators as referred by Eqs. \eqref{qkg_Eq15} \& \eqref{qkg_Eq16}. 
\begin{equation}
e^{-i\hat{H_{1}}t}=F^{-1}e^{- i\hat{p_{p}}^2\Box}.F.F^{-1}.e^{- i\hat{p_{p}}^2}\Box F.e^{-iV_{1}\sigma_{z}t} 
\label{qkg_Eq15}
\end{equation}
\begin{equation}
e^{-i\hat{H_{2}}t}=F^{-1}e^{ i\hat{p_{p}}^2\Box}.F.F^{-1}.e^{ i\hat{p_{p}}^2\Box}F^.e^{-iV_{2}\sigma_{z}t}
\label{qkg_Eq16}
\end{equation}
where $\hat{V_{1}}=\hat{V}+1$ and $\hat{V_{2}}=\hat{V}-1$.
Here we take the high potential of the barrier as $V_{0}$=11 so that the potentials $V_{1}$ and $V_{2}$ are 12 and 10 respectively.

\subsection{Experimental Results}
\begin{figure*}
\begin{center}
\epsfig{file=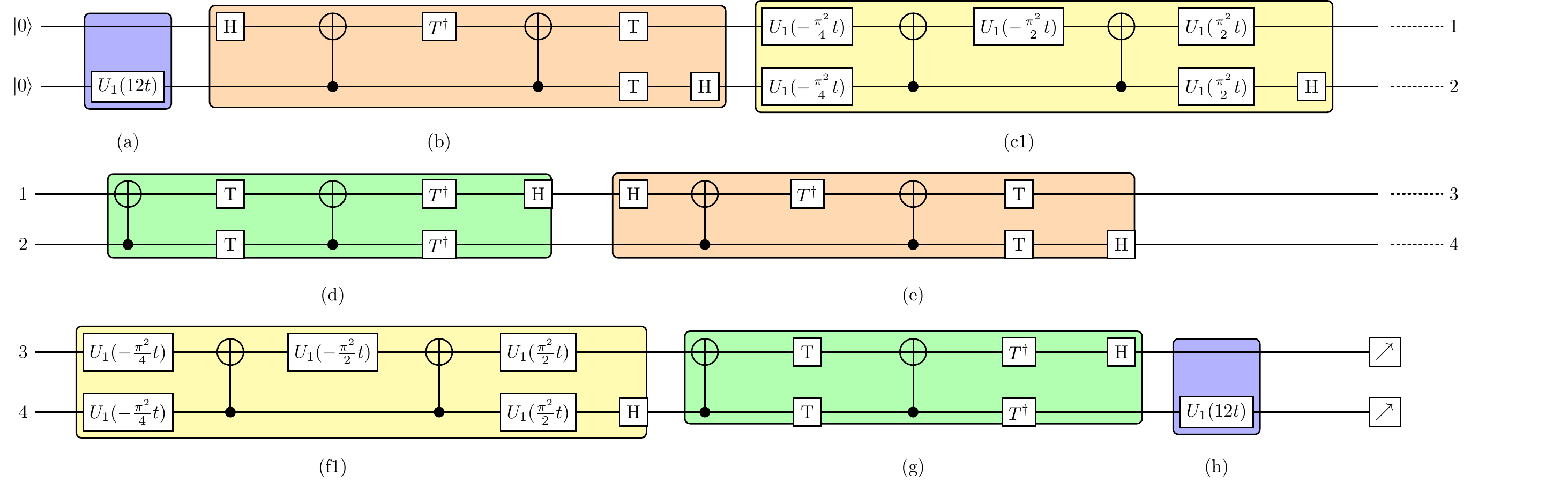,width=16cm}
\epsfig{file=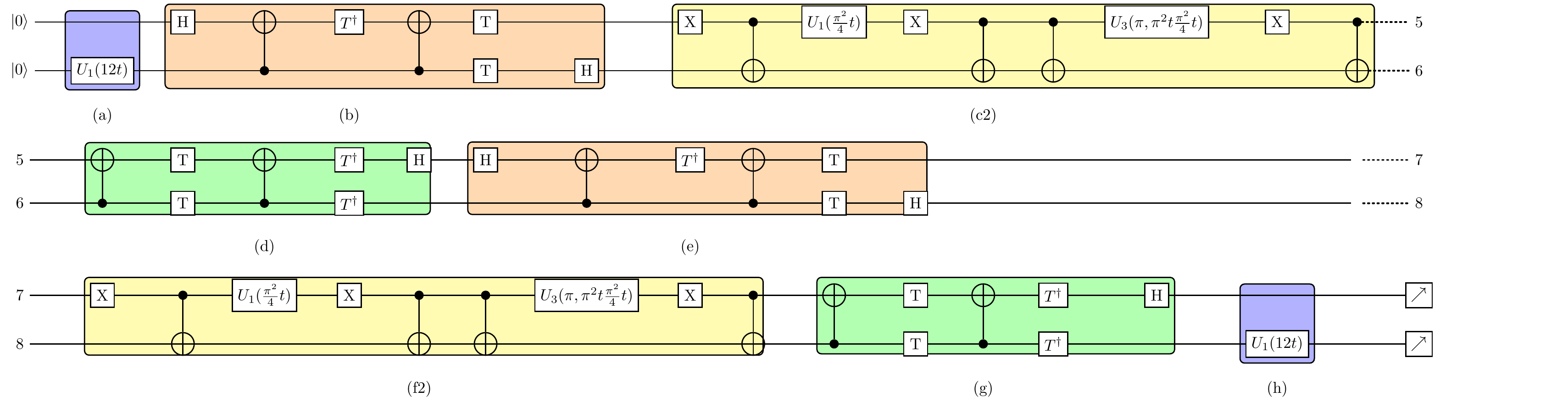,width=16cm}
\caption{\textbf{Quantum circuit for simulating the Klein Gordon equation.} Steps (b) and (d) represent the circuits performing the inverse of the Fourier transform $F^{\dagger}$. Steps (d) and (g) represent the circuit for the Fourier transform F. Steps (a) and (h) represent the potential term of the Klein Gordon equation. Steps (c1), (c2), (f1), and (f2) are the digital quantum circuit representation for the eigenvalues of the respective eigenfunctions.}
\label{qkg_Fig1}
\end{center}
\end{figure*}

\begin{figure*}
\begin{center}
\epsfig{file=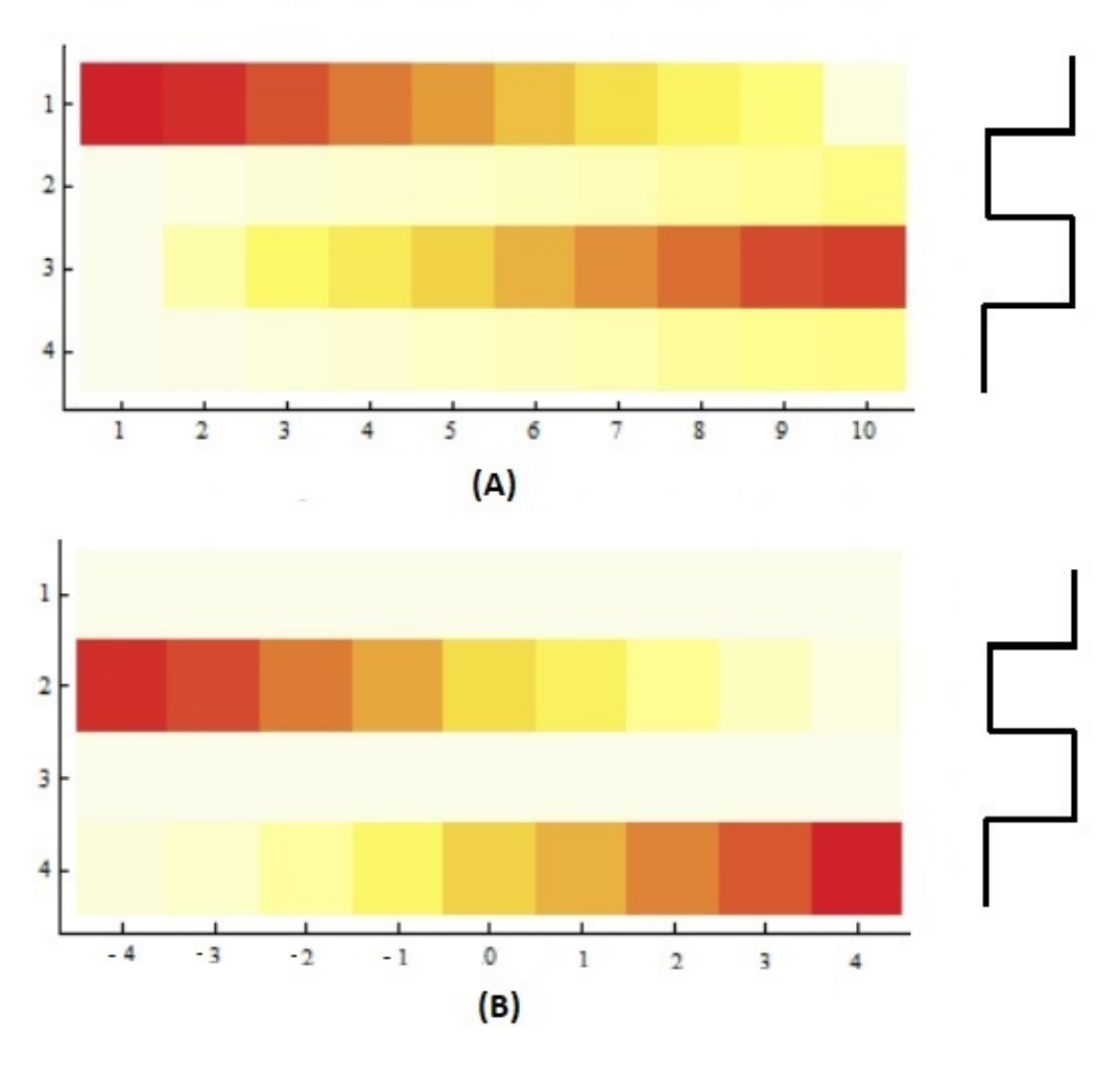,width=16cm}
\caption{\textbf{Experimental results for the two parts of the wavefunction $\varphi$}. Here in the vertical axis 1, 2, 3 and 4 represent $\ket{00}$, $\ket{01}$, $\ket{10}$ and $\ket{11}$ respectively. The horizontal axis represents the time scale of the evolution of the quantum state. As time goes on we can see that the particle slowly crosses over the barrier potential as expected from the condition, $V>mc^2$ (Klein Paradox). \textbf{Case (A):} The particle was initially in the state $|00\rangle$ at time t=1. After 10 iterations at time t=10, it was found in state $|10\rangle$. \textbf{Case (B):} The anti-particle was located in state $|01\rangle$ in the negative time scale. After 8 iterations at time t=4, the anti-particle was found to be located in state $|11\rangle$. In the above cases, we clearly observe the tunneling of both the particle and anti-particle through lower and higher potentials respectively.}
\label{qkg_Fig2}
\end{center}
\end{figure*}
To solve the equation, we use two qubits for representing four lattice points and a small barrier potential for illustrating the tunneling process in the system. The quantum circuits provided in Fig. \eqref{qkg_Fig2}, \textbf{Case (A)} and \textbf{(B)} are used to solve the Eqs. \ref{qkg_Eq15} \& \ref{qkg_Eq16} respectively for the four lattice points. We then run the quantum circuits for different values of t and observe the dynamics of the system by performing a digital simulation. In the circuit, a number of iterations of the circuits were done and a higher amount of iterations of the circuit led to increased accuracy in the results. The number of iterations were kept between 9-10 in most of the cases though even 6 provided appropriate results. Here, t is the only variable in our program which decides the output at various time instances and this output was observed to find the location of the particle at various instances. 

In Fig. \ref{qkg_Fig2}, \textbf{Case (A)} and \textbf{Case (B)} represent the evolution of a quantum system for a particle and an anti-particle respectively as a consequence of Klein Gordon equation. The tunneling is obersved for both the cases as a consequence of Klein Paradox due to a strong potential, i.e., $V > mc^2$, where $V=11$ and $mc^2=1$. In \textbf{Case (A)}, it can be clearly observed that the particle was initially located in $|00\rangle$ state at time, t=1 in a relatively lower potential region as compared to \textbf{Case (B)}. At time t=6, it was in an equal superposition of states $|00\rangle$ and $|10\rangle$. Then after the application of a number iterations, it slowly tunneled through the barrier of the potential which was at location $|01\rangle$ state. After 10 iterations, i.e., at time t=10, the particle was found to completely cross the barrier and finally located in $|10\rangle$ state. In \textbf{Case (B)}, the anti-particle was observed in the negative time scale and in a comparatively higher potential region that clearly indicates its nature that is impossible for a particle. At time, t=-4, it was in the state $|01\rangle$, then it was found in an equal superposition of $|01\rangle$ and $|11\rangle$ states at time t=0. After 8 iterations, i.e., at time t=4, it was noticed to completely tunnel through the potential situated at $|10\rangle$ state, and finally found to be in $|11\rangle$ state. 

\section*{\textbf{Discussion}}
Here we have proposed a two-qubit quantum circuit for the simulation of Klein Gordon equation to study the nature of particle and anti-particle behaviour. Quantum tunneling of both the particle and anti-particle has been observed due to a strong potential as a consequence of Klein Paradox. The simulation technique used here can be extended to higher dimensional lattice structure and using n=$log_{2}N$ number of qubits for N-qubit lattice, the dynamics of the system for any physical system of the equation can be studied. The proposed digital simulation is very much useful for study of a number of quantum field-theoretic equations. 

\section*{\textbf{Methods}}

The Kinetic energy operator plays a key role in our problem and solving it is of paramount importance. We start by taking the Fourier transform of the kinetic energy operator according to the following equation. 
\begin{equation}
e^{\pm i\frac{\hat{p_{x}}^2}{2m}()}=F^{-1}e^{\pm i\frac{\hat{p_{p}}^2}{2m}}.F
\end{equation}
The form of Fourier transform for the two qubits can be written as follows.
\begin{equation*}
F =
\left[{\begin{array}{cccc}
                    1&1&1&1\\
                   1&i&-1&-i\\
                   1&-1&1&-1\\
                   1&-i&-1&i\\
\end{array} } \right] 
\end{equation*}
To find the momentum eigenvalue matrix given by $e^{\pm i\frac{\hat{p_{p}}^2}{2m}}$, we need to first find the eigenstates of the matrix in the $\hat{X}$ coordinate representation. A wavefunction in the terms of momentum eigenvalues can be expressed as
\begin{equation*}
\ket{\phi(p,t)}=\sum_{j=0}^{2^n-1} \phi(p_{j},t)\ket{j}
\end{equation*}
The eigenvalues of this state can be found using a formula
\begin{equation*}
p_{j}=
\begin{cases}
\frac{2\pi}{2^n}j & 0\leqslant j\leqslant2^{n-1}
\\
\frac{2\pi}{2^n}(2^{n-1}-j) & 2^{n-1} < j < 2^{n}
\end{cases}
\end{equation*}
On calculating this we find that the matrix is diagonal and is given by the following expression. 
\begin{equation*}
\hat{P}_{p}=\sum_{j=0}^{2^{n-1}}\frac{2\pi}{2^n}j\ket{j}\bra{j}+\sum_{j=2^{n-1}+1}^{2^n-1}\frac{2\pi}{2^n}(2^{n-1}-j)\ket{j}\bra{j}
\end{equation*}
Solving for two qubit simulation we have
\begin{equation*}
\hat{P}_{p}=
\frac{2\pi}{4}
\left[{\begin{array}{cccc}
                    0&0&0&0\\
                    0&1&0&0\\
                    0&0&2&0\\
                    0&0&0&-1\\
\end{array} } \right] 
\end{equation*}
If we take m=0.5 we get the following equation as our kinetic energy operator
\begin{equation}
e^{\pm i\frac{\hat{p_{x}}^2}{2m}()}=F^{-1}e^{\pm A}.F
\end{equation}
where 
\begin{equation*}
A=
\frac{\pi^2}{4}
\left[{\begin{array}{cccc}
                    0&0&0&0\\
                    0&1&0&0\\
                    0&0&4&0\\
                    0&0&0&1\\
\end{array} } \right] 
\end{equation*}
\section*{References}
\bibliographystyle{naturemag}

\section*{Acknowledgements} M.K. acknowledges the hospitality provided by Indian Institute of Science Education and Research Kolkata. M.K. acknowledges Daattavya Aggarwal for helping with QISKit. B.K.B. acknowledges the financial support of DST Inspire Fellowship. We are extremely grateful to IBM team and IBM Quantum Experience project. The discussions and opinions developed in this paper are only those of the authors and do not reflect the opinions of IBM or IBM QE team. 

\section*{Author contributions}
M.K. has investigated the theoretical analysis, designed the quantum circuit, performed the experiment and taken the data using QISKit. M.K. and B.K.B. have the contribution to the composition of the manuscript. B.K.B. has supervised the project. M.K. and B.K.B. have completed the project under the guidance of P.K.P. 

\section*{Competing interests}
The authors declare no competing financial as well as non-financial interests. 

\end{document}